\documentclass[aps,prl,twocolumn,floatfix,superscriptaddress]{revtex4-1}
\usepackage{amsmath,amsfonts,amssymb}

\usepackage{graphicx}
\usepackage{chngcntr}
\usepackage[caption=false]{subfig}

\newcommand{\bra}[1]{\ensuremath{\left\langle #1\right\vert}}
\newcommand{\ket}[1]{\ensuremath{\left\vert #1\right\rangle}}

\newcommand{\hsp}[1]{\hspace{#1 em}}
\newcommand{\sqz}{\hsp{-0.1}}
\newcommand{\ketbra}[2]{\left\vert{#1}\right\rangle \sqz\sqz\sqz \left\langle{#2}\right\vert}
\newcommand{\braket}[2]{\left\langle{#1}\right\vert \sqz \sqz \left. {#2}\right\rangle}

\newcommand{\iden}{\mathbf{1}}
\newcommand{\tr}[1]{\mathrm{Tr}(#1)}

\newcommand{\myemph}[1]{\textbf{\textit{#1}}}

\begin{document}
\title{Converting non-classicality into entanglement}
\date{\today}

\author{N. Killoran}
\affiliation{Institut f\"{u}r Theoretische Physik, Albert-Einstein-Allee 11, Universit\"{a}t Ulm, D-89069 Ulm, Germany}
\author{F. E. S. Steinhoff}
\affiliation{Naturwissenschaftlich Technische Fakult\"{a}t, Universit\"{a}t Siegen, Walter-Flex-Str. 3, D-57068 Siegen, Germany}
\affiliation{Instituto de F\'isica, Universidade Federal de Goi\'as, 74001-970, Goi\^ania, Goi\'as, Brazil}
\author{M. B. Plenio}
\affiliation{Institut f\"{u}r Theoretische Physik, Albert-Einstein-Allee 11, Universit\"{a}t Ulm, D-89069 Ulm, Germany}

\begin{abstract}
Quantum mechanics exhibits a wide range of non-classical features, of which entanglement in multipartite systems takes a central place. In several specific settings, it is well-known that non-classicality (e.g., squeezing, spin-squeezing, coherence) can be converted into entanglement. In this work, we present a general framework, based on superposition, for structurally connecting and converting non-classicality to entanglement. In addition to capturing the previously known results, this framework also allows us to uncover new entanglement convertibility theorems in two broad scenarios, one which is discrete and one which is continuous. In the discrete setting, the classical states can be any finite linearly independent set. For the continuous setting, the pertinent classical states are `symmetric coherent states,' connected with symmetric representations of the group $SU(K)$. These results generalize and link convertibility properties from the resource theory of coherence, spin coherent states, and optical coherent states, while also revealing important connections between local and non-local pictures of non-classicality.
\end{abstract}

\maketitle


Quantum mechanics currently provides our deepest description of nature. Despite this, much of our everyday experience can be accurately captured within a classical description. What is special about the non-classical states of a physical system, and what distinguishes them from the more commonplace classical states? Certainly, one of the most important manifestations of non-classicality is entanglement of multipartite systems. Schr\"{o}dinger even viewed entanglement as ``\emph{the} characteristic trait of quantum mechanics'' \cite{schrodinger35a}. Yet there are situations where entanglement has no natural role in describing non-classicality, such as Fock states in optics \cite{leonhardt97a}. Particularly for non-composite systems, other notions of non-classicality appear better suited for characterizing quantum states.

A key aspect where quantum mechanics departs from classical mechanics is the prominence of the superposition principle. This elementary tenet of quantum theory supplies a very general framework for categorizing classical and non-classical states. Depending on the particular setting, we may specify some important subset of pure states $\{\ket{c}\}_{c\in\mathcal{I}}$ to be the `classical' pure states of the system. We can then directly associate non-classicality with superposition: a state $\ket{\psi}$ is non-classical if and only if it is a non-trivial superposition of classical states. In fact, entanglement fits naturally within this superposition framework, by specifying factorized states as the classical states. 

One famous example of classical states is that of optical coherent states, $\{\ket{\alpha}\}_{\alpha\in\mathbb{C}}$ \cite{sudarshan63a,glauber63a,leonhardt97a}. A completely different example is found in the resource theories of coherence \cite{baumgratz14a} or reference frames \cite{gour08a}, where the classical states are some fixed orthonormal basis $\{\ket{k}\}_{k=0}^M$. In both examples, there is no distinction of subsystems, and entanglement is not obviously relevant. Nevertheless, there are fundamental connections between these single-system concepts of classicality and the multipartite property of entanglement. It is well-known that a beamsplitter (with the second port in vacuum) transforms optical coherent states as $\ket{\alpha}\otimes\ket{\mathrm{vac}}\rightarrow\ket{r\alpha}\otimes\ket{t\alpha}$. Analogously, a generalized controlled-NOT (with the target in $\ket{0}$) has the effect $\ket{k}\otimes\ket{0}\rightarrow\ket{k}\otimes\ket{k}$. For both cases, the given 
operation transforms all the classical states into factorized states. Importantly, the same operation transforms all non-classical states into entangled states \cite{kim02a, xiang02a, asboth05a, steinhoff12a, jiang13a, vogel14a, streltsov15a}. Put another way, these transformations faithfully convert non-classical resource states into entangled resource states. This connection, illustrated in two quite different settings, provokes intriguing questions. How general is this convertibility property? For a given notion of non-classicality, can we always convert the non-classical states into entangled states, while leaving the classical states unentangled? 

In this Letter, we show that faithful unitary conversion is possible in two wide-ranging new scenarios -- one discrete, one continuous. In the discrete setting, the classical pure states can be an arbitrary linearly independent set. This generalizes the notions of classicality and convertibility from the resource theory of coherence \cite{baumgratz14a, streltsov15a}. For the continuous setting, the applicable classical states are generalized coherent states associated with symmetric representations of the group $SU(K)$, for $2\leq K < \infty$. Such states bridge the gap between the two-level spin coherent states \cite{radcliffe71a, perelomov72a, gilmore72a, arecchi72a} and the infinite dimensional optical coherent states. Furthermore, for both the discrete and continuous scenarios, we outline the 
operations which carry out the desired conversions. These results provide valuable new insights into the resource theory of coherence \cite{baumgratz14a, streltsov15a}, the classical nature of coherent states \cite{perelomov72a,gilmore72a,zhang90a}, the nature of entanglement in identical particles \cite{cavalcanti07a,killoran14a,tasgin15a}, and the basic structure of quantum mechanics. 


\emph{Classical and non-classical ---}
There are various concepts of non-classicality, each relevant to a particular setting. Interestingly, the same state can be seen as classical in one setting and non-classical in another. For instance, excluding the vacuum, Fock states $\{\ket{n}\}_{n=1}^\infty$ can be thought of as non-classical because they are superpositions of `classical' optical coherent states. Alternatively, we can see them as classical, since, by orthogonality, mixtures of these states are in one-to-one correspondence with classical probability distributions. Because these different perspectives each have their uses, we must recognize that non-classicality is a relative notion. For our purposes, we will allow the set of classical pure states to be arbitrarily specified, and assume there is some justification for the choice. Thus, we simply have a list of classical pure states $\mathcal{C}_P:=\{\ket{c}\in \mathcal{H}\}_{c\in\mathcal{I}}$, where $\mathcal{H}$ is a Hilbert space of dimension $D$ and 
$\mathcal{I}$ is some indexing set. We permit an arbitrary number of classical states, even a 
continuous set (e.g., as with coherent states). Unfortunately, in the resource theory of coherence, the term `coherent' applies to  \emph{non-classical} states, while in optics it is used for the \emph{classical} states. To avoid confusion, we will reserve the name `coherent state' for the latter setting and its group-theoretic generalizations.

Suppose that the set $\mathcal{C}_P$ has been specified. To maintain the desired correspondence between non-classicality and superposition, we take $\mathcal{H}$ as the span of the classical states. Every pure state $\ket{\psi}\in\mathcal{H}$ can thus be expanded using some superposition of classical states. We extend our framework to mixed states with one further requirement, namely that convex combinations of classical states are classical \footnote{Note that this does not capture all interesting types of non-classicality. For example, quantum discord does not fit into this formalism, since the set of zero-discord states is non-convex}. Thus the full set of classical states is given by the convex hull of the specified classical pure states, $\mathcal{C}:=\mathrm{conv}(\mathcal{C}_P)$. Any state which cannot be written as a convex combination of classical pure states will be called non-classical, and we denote the set of all such states as $\mathcal{NC}$. 
Together, this partitions the state space into two disjoint sets. When the classical pure states are finite, we can make the following definition. Out of all possible superpositions, there will be some \emph{minimal} number $1\leq r_C\leq D$ of non-zero terms that must be used. We will call this the \emph{C-rank} $r_C$ of $\ket{\psi}$:
\begin{align}
	r_C(\ket{\psi}):=\min \left\lbrace r ~ \Bigg| \ket{\psi}=\sum_{j=1}^r \psi_j\ket{c_j^{(\psi)}} \right\rbrace,
\end{align}
where the states $\ket{c_j^{(\psi)}}$ are each classical (cf. \cite{sperling10a}). All classical states have $r_C=1$ and all non-classical states necessarily have $r_C>1$. Even if the classical states are overcomplete and different decompositions are possible, the C-rank is a well-defined quantity. We point out the conceptual similarity with the Schmidt rank from entanglement theory. For continuous $\mathcal{I}$, one might instead expand a state using an integral over classical states; however, the notion of C-rank for such systems is not so clear. 

We can always convert from a single-system picture to a bipartite picture using the following procedure. The initial system is connected to an ancilla system (with Hilbert space $\mathcal{H}_\mathrm{anc}\cong\mathcal{H}$) which is in a fixed reference classical 
state $\ket{\psi_\mathrm{ref}}$ \footnote{For the present theorems, we do not require $\ket{\psi_\mathrm{ref}}$ to be classical. However, conceptually this is the most sensible choice, especially for resource theory applications.}. We apply some global operation $\Lambda$ to the combined system. Since $\Lambda$ is non-local, it has the potential to create entanglement where none existed before. The entanglement properties of the final state depend on both the chosen global operation and on the input state. The ancilla's role is passive, i.e., it should not contribute anything to the final state's entanglement. The goal is that $\Lambda$ produces an entangled output state if and only if the input state is non-classical. In other words, $\Lambda[\mathcal{C}]\subset \mathcal{S}$ and $\Lambda[\mathcal{NC}]\subset \mathcal{E}$, where $\mathcal{S}$ and $\mathcal{E}$ are, respectively, the separable and entangled states on the output space. A conversion 
$\Lambda$ will be considered faithful when this property holds, since the partitionings on both the input and output spaces are respected \footnote{One could also consider unfaithful conversions, allowing some non-classical states to become separable.}. We can picture the overall protocol not as the creation of entanglement out of nothing, but rather as the conversion of non-classicality into entanglement. It has been recognized previously in setting-specific scenarios \cite{asboth05a, vogel14a, streltsov15a, tasgin15b, miranowicz15a} that non-classicality can be quantified using entanglement measures. Such methods fundamentally require that only non-classical states have the potential to generate entanglement. Complementary results for discord-type quantum correlations have also been developed \cite{streltsov11a,piani11a,gharibian11a,piani12a,nakano13a}. We explore here the qualitative aspects of non-classicality conversion, postponing quantitative questions to future work.

To construct the conversion operations, we will leverage a useful theorem from \cite{chefles04a,marvian13a} which involves Gram matrices. Before stating it, we quickly review a few helpful definitions and properties. For a fixed set of states $\{\ket{\psi_i}\}_{i=1}^N$, we define an $N\times N$ Gram matrix $G^{(\psi)}$ by 
\begin{align}\label{eq:grammatrix}
[G^{(\psi)}]_{ij}=\braket{\psi_i}{\psi_j}.	
\end{align}
For any Gram matrix, we have that $G^{(\psi)}\geq 0$, and $\mathrm{rank}(G^{(\psi)})$ equals the number of linearly independent vectors in $\{\ket{\psi_i}\}_{i=1}^N$. Further, when the states are normalized, $\mathrm{diag}(G^{(\psi)})=\mathrm{diag}(\iden)$, and Gram matrices for product states $\{\ket{\psi_i}\otimes\ket{\phi_i}\}_{i=1}^N$ necessarily have the form $G^{(\psi,\phi)}=G^{(\psi)}\circ G^{(\phi)}$, where `$\circ$' denotes the entrywise Hadamard product $[X\circ Y]_{ij}=X_{ij}Y_{ij}$. Finally, every $N$-dimensional matrix $M\geq 0$ with $\mathrm{diag}(M)=\mathrm{diag}(\iden)$ is the Gram matrix for some appropriate set of states $\{\ket{\delta_i}\}_{i=1}^N$ (determined from the columns of $C$ in $M = C^\dagger C$). If $\{\ket{\psi_i}\}_{i\in\mathcal{I}}$ is a continuous set, we can consider a two variable function $G^{(\psi)}(i,j)=\braket{\psi_i}{\psi_j}$ analogous to Eq. (\ref{eq:grammatrix}), which we will, for convenience, also call a Gram matrix. 

\noindent \myemph{Theorem 1 (Unitary conversion) \cite{chefles04a, marvian13a}:} Let $\{\ket{\psi_i}\}_{i\in \mathcal{I}}$ and $\{\ket{\phi_i}\}_{i\in \mathcal{I}}$ be two sets of states. There exists a unitary operation $\Lambda$ such that $\Lambda\ket{\psi_i}=\ket{\phi_i}$ for all $i\in\mathcal{I}$ if and only if $G^{(\psi)}=G^{(\phi)}$ \footnote{Note that \cite{chefles04a} allows more flexibility with global phases. We will not need this extra detail.}.

\pagebreak

\emph{Discrete case ---} In the discrete setting, we fix the set of classical pure states to be finite, $\mathcal{C}_P=\{\ket{c_i}\}_{i=1}^D$. We can immediately state our first main result.

\noindent \myemph{Theorem 2 (Discrete convertibility):} If the classical pure states $\{\ket{c_i}\}_{i=1}^D$ are linearly independent, then there exists a unitary $\Lambda$ such that, for all $\ket{\psi}\in\mathcal{H}$, the Schmidt rank of $\Lambda\ket{\psi}$ is equal to the C-rank of $\ket{\psi}$. For mixed states, we have $\Lambda\rho\Lambda^\dagger\in\mathcal{S}$ if and only if $\rho\in\mathcal{C}$.

\emph{Proof:} If $\{\ket{c_i}\}_{i=1}^D$ are linearly independent, then $G^{(c)}$ is full rank and hence $G^{(c)}>0$. Using a construction of \cite{djokovic65a}, define a $D\times D$ matrix $B(\lambda)$ with entries $B_{ij}=\lambda$ for $i\neq j$ and $\mathrm{diag}(B)=\mathrm{diag}(\iden)$. For $0\leq\lambda <1$, we have $B(\lambda)>0$. The matrix $M(\varepsilon) := G^{(c)}\circ B(1+\varepsilon)$ is Hermitian and $M(\varepsilon)> 0$ for sufficiently small $\varepsilon > 0$ since $\lim_{\varepsilon\rightarrow 0^+}M(\varepsilon)=G^{(\alpha)}>0$. Choosing any valid $\varepsilon$, we have $G^{(c)}=B(\tfrac{1}{1+\varepsilon})\circ M(\varepsilon)$, with $B(\tfrac{1}{1+\varepsilon})>0$ and $M(\varepsilon)>0$. In fact, both $B(\tfrac{1}{1+\varepsilon})$ and $M(\varepsilon)$ have only ones on their diagonals, so we actually have $B(\tfrac{1}{1+\varepsilon})=G^{(d)}$, $M(\varepsilon)=G^{(e)}$ where $G^{(d)}$ and $G^{(e)}$ are Gram matrices for some linearly independent sets $\{\ket{d_i}\}_{i=1}^D$ and 
$\{\ket{e_i}\}_{i=1}^D$. 

From the above properties, $G^{(c)}=G^{(d)}\circ G^{(e)}$ is the Gram matrix for the product states $\{\ket{d_i}\otimes\ket{e_i}\}_{i=1}^D$. Therefore, there exists a unitary $\Lambda$ such that $\Lambda\ket{c_i}=\ket{d_i}\otimes\ket{e_i}~\forall~i=1,\dots,D$. Finally, let $\ket{\psi}\in\mathcal{H}$ have C-rank $r_C$. Then $\ket{\psi}=\sum_{j=1}^{r_C}\psi_j\ket{c_{\pi(j)}}$ where $\pi$ is some permutation of $\{1,\dots,D\}$ depending on $\psi$. Thus, $\Lambda\ket{\psi}=\sum_{j=1}^{r_C} \psi_j\ket{d_{\pi(j)}}\otimes \ket{e_{\pi(j)}}$. Because the states $\{\ket{d_{\pi(j)}}\}_{j=1}^{r_C}$ and $\{\ket{e_{\pi(j)}}\}_{j=1}^{r_C}$ are locally linearly independent, it follows that the Schmidt rank of the output state $\Lambda\ket{\psi}$ will be exactly $r_C$. For mixed states, it is easily checked that $\rho\in\mathcal{C}\Rightarrow\Lambda\rho\Lambda^\dagger\in\mathcal{S}$. Conversely, observe that the only factorized (i.e., Schmidt rank 1) states in the 
image $\Lambda[\mathcal{C}_P]$ are exactly the states $\{\ket{d_i}\otimes\ket{e_i}\}_{i=1}^D$. Thus, we can conclude that $\Lambda\rho\Lambda^\dagger\in\mathcal{S}\Rightarrow\rho\in\mathcal{C}$. 
\hfill $\square$

Any valid $\Lambda$ from Theorem 2 is a faithful non-classicality to entanglement conversion operation. Since $\mathrm{span}(\{\ket{c_i}\})=\mathcal{H}$, a particular transformation $\Lambda$ is completely specified by the corresponding Gram matrices, and hence by the continuous parameter $\varepsilon$. The infinitely many possibilities correspond to different possible ways of splitting the initial overlap structure between the two new subsystems. The splitting procedure can even be iterated to give multipartite output states. Of course, for more than two subsystems, one would have to consider generalizations of the Schmidt decomposition. We note that our framework also permits splitting of the overlaps in other (non-equally weighted) ways, though these may be more dependent on the particular classical states. We provide an example application of Theorem 2 in section S1B of the Supplemental Material (SM).

In the resource theory of coherence, the classical pure states are orthogonal. By Theorem 2, any superposition of these can be faithfully converted into an entangled state. A prototypical conversion operation is controlled-displacement, which takes 
$\ket{k}\otimes\ket{0}\rightarrow\ket{k}\otimes\ket{k}$ \cite{steinhoff12a,streltsov15a}. This transformation arises in our proof in the limit $\varepsilon\rightarrow\infty$. But our result also applies to a more general notion of coherence, where the classical states are not orthogonal. Interestingly, the conversion transformations are close analogs: instead of controlled-displacements, we have $\ket{c_k}\otimes\ket{c_0}\rightarrow\ket{d_k}\otimes\ket{e_k}$. At present, less is known about this more general notion of non-classicality. Non-orthogonal states are important in quantum foundations \cite{pusey12a}, quantum key distribution \cite{bennett84a}, and quantum state estimation \cite{chefles98a}. However, an abstract framework for linear independence, similar to the resource theories of coherence or entanglement, has not to our knowledge been constructed. Nevertheless, we now know that this form of non-classicality is intimately connected to entanglement. 


\emph{Continuous case ---} In the introduction, we identified another notion of classical states: the \emph{optical coherent states} $\{\ket{\alpha}\}_{\alpha\in\mathbb{C}}$. Splitting these states is accomplished using a beamsplitter (parameterized by $(r,t)$, with $|r|^2+|t|^2=1$). In the Gram matrix formalism, we have
\begin{align}\label{eq:opcohsplit}
	\braket{\alpha}{\beta} 
	& = \braket{\alpha}{\beta}^{|r|^2}\braket{\alpha}{\beta}^{|t|^2}
	= \braket{r\alpha}{r\beta}\braket{t\alpha}{t\beta}.
\end{align}
Optical coherent states are strongly connected to representations of the \emph{Heisenberg-Weyl group} \cite{zhang90a}.
The states $\ket{r\alpha},\ket{t\alpha}$ can be thought of as belonging to separate `rescaled' representations of the coherent states, 
with displacement operators $\hat{D}_r(\alpha):=\exp(r\alpha\hat{a}^\dagger - h.c.)$. We can thus view a beamsplitter as a physical operation that transforms between different (bipartite) representations of the 
coherent states, while preserving the underlying group structure (mathematically, this is called an \emph{equivariant map} or \emph{intertwiner} \cite{fecko06a}). 

In fact, generalized coherent states can be constructed for any group \cite{perelomov72a,gilmore72a,zhang90a}. We need the following ingredients: i) an abstract group $G$; ii) an irreducible representation (irrep) of the group as unitary operators $\hat{D}_{q}(g)$ on a Hilbert space $\mathcal{H}$ (where $q$ labels the particular irrep); and iii) a reference state $\ket{\Phi_0}\in\mathcal{H}$. The group coherent states are given by the set 
\begin{align}
\{\ket{g;q} := \hat{D}_q(g)\ket{\Phi_0} ~ | ~ g\in G\},	
\end{align} 
where states differing by a global phase are considered equivalent. An alternate way to generalize coherent states is explored in section S4 of the SM.

We consider the group $SU(K)$, for arbitrary $2\leq K < \infty$, i.e., all possible unitary transformations on a $K$-level system. The irreps of $SU(K)$ are strongly connected with permutation symmetry, coming in symmetric, antisymmetric, and mixed symmetry types (see, e.g., \cite{eichmann13a}). Our results focus on the symmetric irreps (labeled by natural numbers $N$), where an element $U\in SU(K)$ is represented as the unitary operator $\hat{D}_{N}(U):=U^{\otimes N}$. With the reference state $\ket{0}^{\otimes N}$, our coherent states take the form $\ket{U;N}:=[{U}\ket{0}]^{\otimes N}$. Importantly, the representing Hilbert space, $\mathcal{H}_{SU(K);N}:=\mathrm{span}(\{\ket{U;N}\})$, is also the symmetric subspace of the larger space $\otimes_{p=1}^N\mathbb{C}^{K}$ \cite{harrow13a}, so its vectors are invariant under any permutation of the 
label $p$. Bosonic particles (and quasiparticles) have such permutation symmetry, but the 
symmetric subspace is also important for state estimation, optimal cloning, and the de Finetti theorem \cite{harrow13a}. For convenience, we refer to $\{\ket{U;N}\}$ as \emph{symmetric coherent states}.

Symmetric coherent states also have the structure of Eq. (\ref{eq:opcohsplit}), except with natural number labels $N_X+N_Y=N$:
\begin{align}\label{eq:boscohsplit}
	\braket{U;N}{V;N} 
	& = \bra{0}U^\dagger V\ket{0}^{N_X}\bra{0}U^\dagger V\ket{0}^{N_Y}\nonumber\\
	& = \braket{U;{N_X}}{V;{N_X}}\braket{U;{N_Y}}{V;{N_Y}}.
\end{align}
Using this property, we can give our next main result.

\myemph{Theorem 3 (Continuous convertibility):} Let $\mathcal{C}_P$ be the symmetric coherent states for some fixed $2\leq K< \infty$ and $2\leq N <\infty$. For every pair of positive integers $(N_X,N_Y)$ with $N_X+N_Y=N$, there is a unitary $\Lambda$ such that $\Lambda\ket{U;N}=\ket{U;{N_X}}\otimes\ket{U;{N_Y}}$ for all $U\in SU(K)$. For mixed states, we have $\Lambda\rho\Lambda^\dagger\in\mathcal{S}$ if and only if $\rho\in\mathcal{C}$.

\emph{Proof:} Consider the Gram matrix of the coherent states $\{\ket{U;N}\}$, denoted by $G^{(N)}(U,V)$. From Eq. (\ref{eq:boscohsplit}), $G^{(N)}(U,V)=G^{(N_X,N_Y)}(U,V)$ for all positive integers $(N_X,N_Y)$ such that $N_X+N_Y=N$ and $\forall ~ U,V\in SU(K)$. Here, $G^{(N_X,N_Y)}$ is the Gram matrix of the set $\{\ket{U;N_X}\otimes\ket{U;N_Y}\}$. Fix any valid pair $(N_X,N_Y)$. By Theorem 1, there exists a unitary $\Lambda$ such that $\Lambda\ket{U;N}=\ket{U;{N_X}}\otimes\ket{U;{N_Y}}$, independent of $U$. Denote the output spaces $\mathcal{H}_{N_{X/Y}}:=(\mathbb{C}^K)^{\otimes N_{X/Y}}$ and let $\ket{\Omega}=\ket{\Omega_{N_X}}\otimes\ket{\Omega_{N_Y}}$, with $\ket{\Omega_{N_{X/Y}}}\in\mathcal{H}_{N_{X/Y}}$, be any factorized state in the image of $\Lambda$. From permutation symmetry, it must have the form $\ket{\Omega}=\ket{\omega}^{\otimes N}$ for some $\ket{\omega}\in\mathbb{C}^K$ \cite{ichikawa08a, wei10a}. Taking $U_\omega\in SU(K)$ where $U_\omega\ket{0}
=\ket{\omega}$, 
we have $\ket{\Omega}=\ket{U_\omega;N_X}\otimes\ket{U_\omega;N_Y}=\Lambda\ket{U_\omega;N}$. For mixed states, clearly $\rho\in\mathcal{C}\Rightarrow\Lambda\rho\Lambda^\dagger\in\mathcal{S}$. Conversely, let $\sigma=\Lambda\rho\Lambda^\dagger$ be separable with respect to $\mathcal{H}_{N_{X}}\otimes \mathcal{H}_{N_{Y}}$. We expand 
$\sigma=\sum_{k} p_k \ketbra{\Omega^k}{\Omega^k}$ 
where $\ket{\Omega^k} =\ket{\Omega^k_{N_X}}\otimes\ket{\Omega^k_{N_Y}}$. Each term in this mixture must be supported only on the symmetric subspace (otherwise $\sigma$ wouldn't be), so by the above argument,  $\ket{\Omega^k}=\ket{U_{\omega_k};N_X}\otimes\ket{U_{\omega_k};N_Y}=\Lambda\ket{U_{\omega_k};N}$ for some $U_{\omega_k}\in SU(K)$. Thus $\sigma=\Lambda\rho\Lambda^\dagger$ for classical $\rho:=\sum_k p_k\ketbra{U_{\omega_k};N}{U_{\omega_k};N}$, and hence $\Lambda\rho\Lambda^\dagger\in \mathcal{S}\Rightarrow \rho \in \mathcal{C}$. \hfill $\square$

As earlier, we can picture the conversion in Theorem 3 as the bipartite transformation $\ket{U;N}\otimes\ket{\mathrm{vac}}\rightarrow\ket{U;{N_X}}\otimes\ket{U;{N_Y}}$, where the reference state $\ket{\mathrm{vac}}$ is the vacuum state. At first, the existence of this transformation could seem obvious, since we explicitly defined symmetric coherent states with a factorized structure. However, it is important to recognize that the \emph{physical encoding} of the coherent states may change during conversion. In particular, we can convert from a setting where the `subsystems' defined by the tensor product are inaccessible into one where they are acessible. This issue of inaccessible subsystems is encountered frequently with systems of identical bosons. In the SM, we present in detail a conversion example connected to this setting.


\emph{Other implications ---} The most direct implication of the above results is to suggest new methods and resources for the physical generation of entanglement. Beyond this, because our non-classicality framework and associated convertibility theorems are quite general, they also lead to a variety of other interesting consequences. We give here a broad overview of these; interested readers can find technical details in the SM. 

First, knowing that non-classicality and entanglement are so closely related allows us to import and export theoretical concepts and tools between the two pictures. For example, given a conversion operator $\Lambda$ and an entanglement witness $W$, we can define a non-classicality witness $\widetilde{W}$ by inverting the conversion and dropping the ancilla, $\widetilde{W}=(\iden\otimes\bra{\psi_\mathrm{ref}})\Lambda^\dagger W\Lambda(\iden\otimes\ket{\psi_\mathrm{ref}})$. If $W$ detects entanglement after conversion, then $\widetilde{W}$ detects non-classicality without needing to convert. Entanglement conversion can also enhance our capabilities in settings where constraints or superselection rules limit our available measurements. For example, in condensed spin systems, we are limited to only collective observables, e.g., the total spin operators $\hat{J}_{k}$. But any single-mode non-classical state (e.g., a spin-squeezed state) can be faithfully converted into its equivalent two-mode entangled form. The bipartite setting then allows us to break the collective symmetry, and measure spin operators on separate subcomponents, $\hat{J}^A_{k}$, $\hat{J}^B_{k}$, in addition to the total system. This extra measurement information can help us detect more non-classicality than in the original setting. These ideas are laid out in more detail in section S2 of the SM.

Another advantage of the general non-classicality framework is that it suggests connections between seemingly unrelated physical settings. On the face of it, the resource theory of coherence and the setting of quantum optics are quite different, since their classical states are completely orthogonal and nonorthogonal, respectively. However, any finite collection of optical coherent states $\{\ket{\alpha_i}\}_{i=1}^N$ is linearly independent \cite{vogel14a}. These states can therefore be split not only using a beamsplitter, but also using the methods of Theorem 2, with any nontrivial superposition becoming entangled. Thus, the notion of non-classicality based on linear independence provides a kind of intermediary setting between its counterparts in the resource theory of coherence and quantum optics. A more specific example of this connection is presented in section S3 of the SM.


\emph{Conclusion ---} Observing that many distinct physical settings share similar fundamental structures, we investigated the question of when single-system non-classicality can be faithfully converted to entanglement. We introduced a general Gram matrix framework which provides a platform linking all previous setting-specific results. Further, we prove that entanglement conversion is possible in two broad new scenarios. Though convertibility is now established in a wide variety of settings, we still do not have a set of universal necessary and sufficient conditions for it. Our results suggest that superposition, long known as a distinguishing feature of quantum mechanics, may be the underlying ingredient connecting quantum resources in so many seemingly different physical settings. 

\begin{acknowledgments}
\emph{Acknowledgments ---}
We thank Oliver Marty and Tobias Moroder for many helpful discussions and Ali Asadian, Otfried G\"{u}hne and Costantino Budroni for comments. This research was supported by an Alexander von Humboldt Professorship, the EU Integrating project SIQS, 
the program Science without Borders from the Brazilian agency CAPES, the EU (Marie Curie CIG 293993/ENFOQI), the BMBF  (Chist-Era  Project  QUASAR),  the  FQXi  Fund (Silicon Valley Community Foundation), the DFG, and CNPq (PVE/PDJ-150733/2015-1).
\end{acknowledgments}

\clearpage
\setcounter{equation}{0}\renewcommand{\theequation}{S\arabic{equation}}


\begin{center}
\textbf{\large Supplemental Material: Converting Nonclassicality into Entanglement}
\end{center}

In the Supplemental Material, we provide more details about the implications and applications of nonclassicality convertibility given in the main text. Specifically, we cover: methods for creating entanglement; the cross-pollination of tools and ideas from entanglement and nonclassicality; a relation between conversion operators in the discrete setting and beamsplitters; and an alternate notion of generalized coherent states. 

\section{S1. Methods for entanglement creation}

The most direct implication of the convertibility results is for physically creating entanglement. We will now go through two specific cases, each connected to one of the convertibility theorems, in more detail.

\subsection{A. Continuous systems: mode splitting}\label{ssec:modesplit}

A common physical setting for symmetric coherent states involves identical bosons in the same spatial mode, such as Bose-Einstein condensates in a harmonic trap, optical lattice, or atom chip, or photons in free space, a fibre, or a waveguide. Suppose we have $N$ particles (or even pseudo-particles), each with $K$ internal levels $\{\ket{j}\}_{j=1}^{K-1}$. For massive particles, these could be ground/excited states; for photons, we could use polarization. A single particle's internal state is represented in a $K$-dimensional Hilbert space, $\mathcal{H}_\mathrm{1}=\mathbb{C}^K$. If the particle is in spatial mode $M$ and internal level $j$, we will denote its state as $\ket{j_M}$. All other internal states will be labelled using unitaries from $SU(K)$, $U\ket{0_M}=:\ket{U_M}$. For a composite system of $N$ particles in the same mode, we can represent the state using the space $\otimes_{p=1}^N\mathbb{C}^K$. The tensor label $p$ indexes the individual single-particle spaces. 

Bosonic exchange symmetries restrict the state of this system to be permutation symmetric, so our overall state space is the symmetric subspace, $\mathcal{H}_{N}:=\mathrm{Sym}[\otimes_{p=1}^N\mathbb{C}^K]$ \cite{harrow13a}. States in the symmetric subspace have the form 
\begin{equation}
\ket{\psi_\mathrm{sym}}=\frac{1}{\sqrt{\mathcal{N}}}\sum_{\mathrm{permutations}~\pi}\otimes_{p=1}^N\ket{\psi_{\pi(p)}},
\end{equation}
where $\ket{\psi_q}$, $q\in\{1,\dots,N\}$ are arbitrary single particle states, $\mathcal{N}$ is some appropriate normalization, and where we sum over all permutations of $N$ elements. When each individual particle occupies the same state, $\ket{\psi_q}=\ket{\psi_{q'}}$, then the corresonding symmetrized state is \emph{algebraically  separable}. Conversely, if any two single-particle states are different, the symmetrized state is \emph{algebraically entangled}. However, \emph{physical entanglement}, i.e., the kind encountered in a resource theory picture with physically separate subsystems, must be interpreted very carefully in this setting. The index $p$, inherited from the single particle descriptions, still serves as a mathematical index for the state spaces in the decomposition. However, in the case of physical bosons, we should not think of $p$ as labeling a physically accessible subsystem. Our particles are completely identical; there is no way to act on `particle $p$' individually (for quasiparticles, it may be possible to break this symmetry). Thus, it is perhaps better in this setting to picture things in terms of \emph{classicality vs. nonclassicality} rather than algebraic separability vs. entanglement.

It is convenient to consider a \emph{second quantization} description for such systems (the above description is called \emph{first quantization}). A symmetric state of $N$ particles, all in the same internal state $U\ket{0}$ and the same spatial mode $M$, will be denoted by $\ket{U;N}_M$. This can be connected with first quantization using the obvious relation 
\begin{equation}\label{eq:classical_spin_states}
	\ket{U;N}_M=\otimes_{p=1}^N\ket{U_M}. 	
\end{equation}
Unlike the index $p$, the mode subscripts $M$ in Eq. (\ref{eq:classical_spin_states}) can refer to physically distinct and individually accessible subsystems; in this case, any entanglement between the modes is physical entanglement. The states $\{\ket{U;N}_M~|~U\in SU(K)\}$ perfectly fit the classicality criteria from Theorem 3. Thus, we know that there is a unitary conversion operator $\Lambda$ which faithfully takes these states to separable states, while converting all nontrivial superpositions into (fully physical) entangled states. In essence, we  convert a system with purely algebraic entanglement (between the identical particles) to one with physical entanglement (between independent modes).

Theorem 3 supplies sufficient information (the action of $\Lambda$ on the basis of classical states) to fully specify the conversion operator. However, one might ask: how can we realize this conversion experimentally? Below, we will outline a simple protocol, requiring only a beamsplitter/tunneling and the ability to count particles, for realizing a specified conversion $\Lambda$ (this is but one approach). The protocol is quite general, and can be applied in many physical settings, for both massive and massless bosons. 

We first need to extend our system to two modes, $A$ and $B$. These are spatially separated, so that the associated single particle states are orthogonal, $\braket{U_A}{V_B}=0$ $\forall~U,V\in SU(K)$. For the moment, consider a single particle, in internal state $U$, but with spatial component $M$ distributed between $A$ and $B$. This particle's state is 
\begin{equation}\label{eq:UM}
	\ket{U_M}=r\ket{U_A}+t\ket{U_B},
\end{equation}
where $|r|^2+|t|^2=1$. If we had $N$ particles in this same state, we can express the global state of the system by
\begin{align}
	\ket{U;N}_M 
	= & \otimes_{p=1}^{N}[r\ket{U_A}_p+t\ket{U_B}_p]\label{eq:tensor1}\\
	= & \sum_{N_A+N_B=N}\hspace{-0.4cm}C_{N_A,N_B}\ket{U; N_A}_A\otimes\ket{U; N_B}_B
	\label{eq:tensor2}
\end{align}
with $C_{N_A,N_B}:=\sqrt{\binom{N}{N_A}}r^{N_A}t^{N_B}$. It is very important to recognize that the tensor products in Eqs. (\ref{eq:tensor1})-(\ref{eq:tensor2}) refer to two different factorizations of the state space. The first is with respect to \emph{particles}, the second is with respect to \emph{modes}. For $|r|\neq 0,1$, a state which is factorized in first quantization is entangled in second quantization, independent of the internal state. This is due to the indefinite particle number
in each mode. 

We now give a way to realize a conversion $\Lambda$. The trick is to actually perform a (conceptually and operationally) simpler unitary, then use a measurement to realize the given $\Lambda$. To be specific, suppose we want to achieve the unitary $\Lambda=\Lambda_{N_X,N_Y}$ which maps the classical states $\{\ket{U;N}\}$ to $\{\ket{U;N_X}_A\otimes\ket{U;N_Y}_B\}$, with ($N_X$,$N_Y$) some fixed particle numbers. Initially, all particles are in mode $A$, and the system is in some superposition
\begin{equation}\label{eq:inputstate}
	\ket{\psi_\mathrm{in}}=\int  \psi_U \ket{U;N}_A d\mu_U,
\end{equation}
where $d\mu_U$ is the natural (Haar) measure for the group $SU(K)$. The mode $B$ is initially independent of $A$ and empty.

For the first stage of the conversion, we introduce an interaction between the two modes via the Hamiltonian 
\begin{equation}
	H_I = \sum_{j=0}^{K-1} e^{i\phi}\hat{a}_j^\dagger \hat{b}_j+e^{-i\phi}\hat{a}_j \hat{b}_j^\dagger
\end{equation}
where $\hat{a}_j^\dagger/\hat{b}_j^\dagger$ are the particle creation operators for internal level $\ket{j}$ and modes $A/B$.
Depending on the experimental 
setting \cite{kim02a,schumm05a,esteve08a,riedel10a, serafini09a}, this is usually called a tunneling or a beamsplitter interaction. The sum over $j$ indicates that the tunneling is agnostic to the internal states of the particles. The modes are allowed to interact for some set time $\tau$, then the interaction is stopped. This can be achieved, for instance, by allowing the spatial domain of modes A and B to partially or fully overlap during the interaction period. The overall transformation is thus given by the unitary
\begin{equation}\label{eq:TrtAB}
	T_{r,t}^{AB} = \exp(i H_I \tau/\hbar).
\end{equation}
In terms of the creation operators, this interaction gives
\begin{align}
	\begin{pmatrix}
		\hat{a}_{i}^\dagger \\
		\hat{b}_{i}^\dagger 
	\end{pmatrix}
	\mapsto
	W_{r,t}^{AB}
	\begin{pmatrix}
		\hat{a}_{i}^\dagger \\
		\hat{b}_{i}^\dagger 
	\end{pmatrix};
	& \hspace{\baselineskip}
	W_{r,t}^{AB}:=
	\begin{pmatrix}
		r & t \\
		t^* & -r^* 
	\end{pmatrix}
\end{align}
for some $r,t$ such that $|r|^2+|t|^2=1$. We assume the interaction is set up so that $|r|\neq 0,1$, i.e., there is a nonzero probability of finding particles in each mode. In first quantization, Eq. (\ref{eq:TrtAB}) can be expressed as the collective unitary $T^{AB}_{r,t}= \otimes_{p=1}^{N} \widetilde{W}^{AB}_{r,t}$, where

\begin{align}
	\widetilde{W}^{AB}_{r,t}\ket{U_A} = & r\ket{U_A}+t\ket{U_B}~\forall~U.	
\end{align}

The classical states $\ket{U;N}_A$ are thus transformed to states matching Eqs. (\ref{eq:tensor1})-(\ref{eq:tensor2}), and an arbitrary input state is transformed to the two-mode output state 
\begin{equation}\label{eq:psistage1}
	\ket{\psi^{r,t}_\mathrm{out}}=\sum_{N_A+N_B=N}\hspace{-0.4cm}C_{N_A,N_B} \int \psi_U \ket{U; N_A}_A\otimes\ket{U; N_B}_B d\mu_U.
\end{equation}
This almost achieves the desired conversion, except for the sum. We supplement it with 
a measurement of local particle numbers. In some experimental
settings, this measurement might be quite difficult; however, it is always physically allowed. The measurement is achieved with the projectors
\begin{equation}
	\{\Pi_{N_1}^A\otimes \Pi_{N_2}^B: N_1+N_2=N\}. 
\end{equation}
Note that the projectors $\Pi_{N_Z}^M$ and $\Pi_{N_Z'}^M$ are orthogonal when $N_Z\neq N_Z'$. Selecting the measurement result ($N_X,N_Y$), the combined operation $\Pi^A_{N_X}\otimes\Pi^B_{N_Y} \cdot T^{AB}_{r,t}$ realizes the desired conversion $\Lambda_{N_X,N_Y}$:
\begin{align}\label{eq:psiout}
	\ket{\psi_\mathrm{out}}= &\int \psi_U \ket{U; N_X}_A\otimes\ket{U; N_Y}_B d\mu_U\nonumber\\
	= & \Lambda_{N_X,N_Y}\ket{\psi_\mathrm{in}}.
\end{align}	
This passive two-stage was discovered previously for the $K=2$ case, and
called \emph{mode-splitting} \cite{killoran14a}.

Measurements are stochastic, so we would only expect to get our desired outcome with probability $|C_{N_X,N_Y}|^2$. However, the underlying group symmetry structure (and hence classicality structure) has not been disturbed by anything we have done (we have just changed our representation from a single-system setting to a bipartite one). Because of this, we can actually repeat the above operations, analogous to \cite{killoran14a}, until the desired outcome is realized. Specifically, if we are seeking the outcome ($N_X,N_Y$), but instead counted ($N_{X'},N_{Y'}$), we simply repeat the protocol, but starting with the state left over from the previous iteration. This can be iterated until the desired outcome is obtained. Of course, in some settings it may be possible to directly apply the unitary $\Lambda_{N_X,N_Y}$; in those cases, the above protocol is unnecessary. Finally, we note that mode-splitting works equally well for generating entanglement out of antisymmetric (fermionic) or even mixed-symmetry coherent states of $SU(K)$. The key difference is that even coherent input states can generate entanglement in those settings, losing the faithful connection between nonclassicality and entanglement.

\subsection{B. Discrete setting: GCNOT}

Suppose we have a qubit system where we cannot create any reliable coherence, so we are restricted to mixtures of the states $\{\ket{0},\ket{1}\}$. Without coherence, the standard entangling gate, the CNOT, is useless for creating entanglement. As well, all transformations of the form $(U^\dagger\otimes\iden) [CNOT] (U\otimes\iden)$, where $U$ creates coherence, are ruled out. Naively, we might conclude that our setup has no possibility for creating entanglement.

\begin{figure}[!t]
    \subfloat[]{%
    \raisebox{8mm}{\includegraphics[width=0.3\columnwidth]{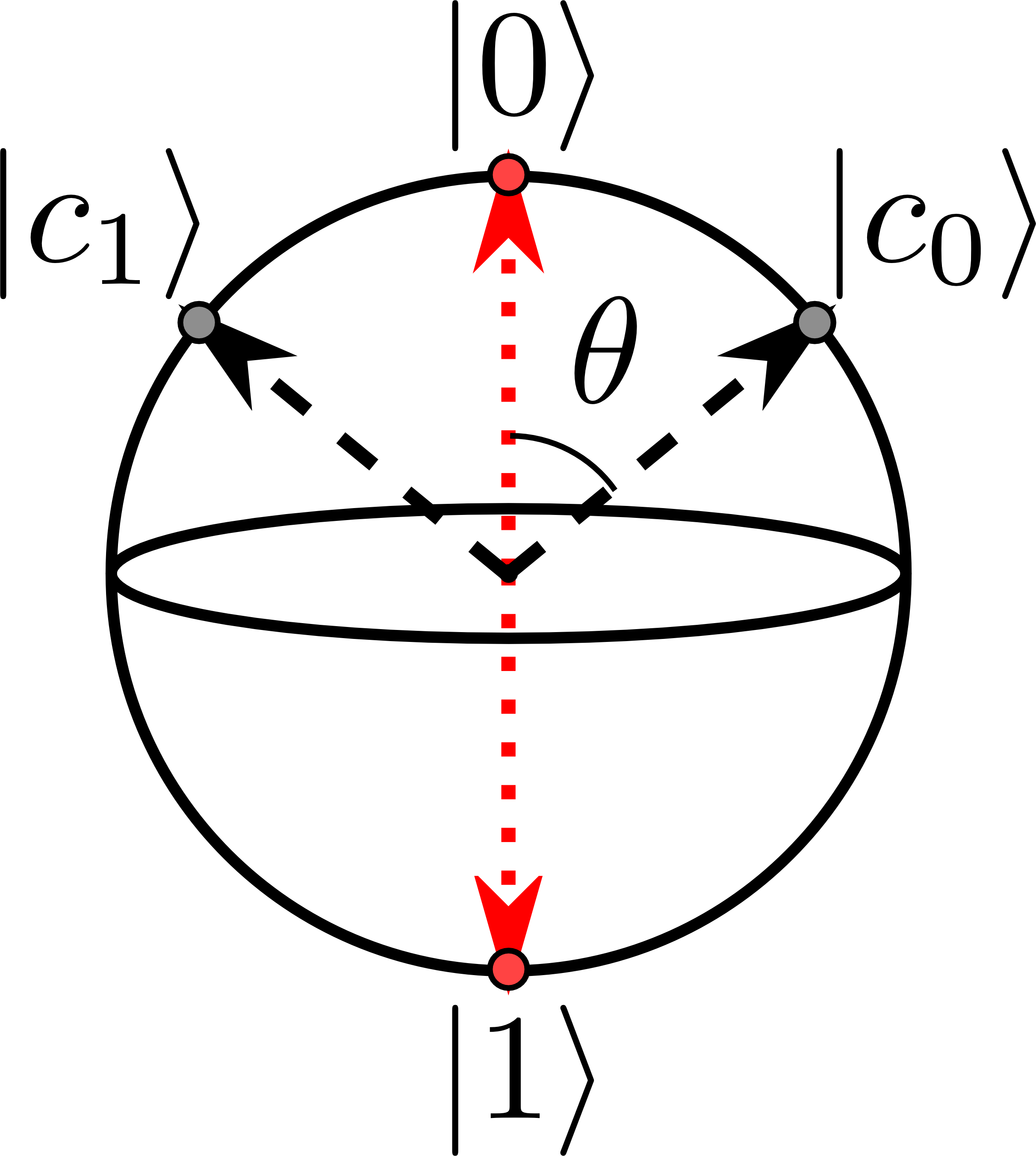}}
    }
    \hfill
    \subfloat[]{%
	\includegraphics[width=0.65\columnwidth]{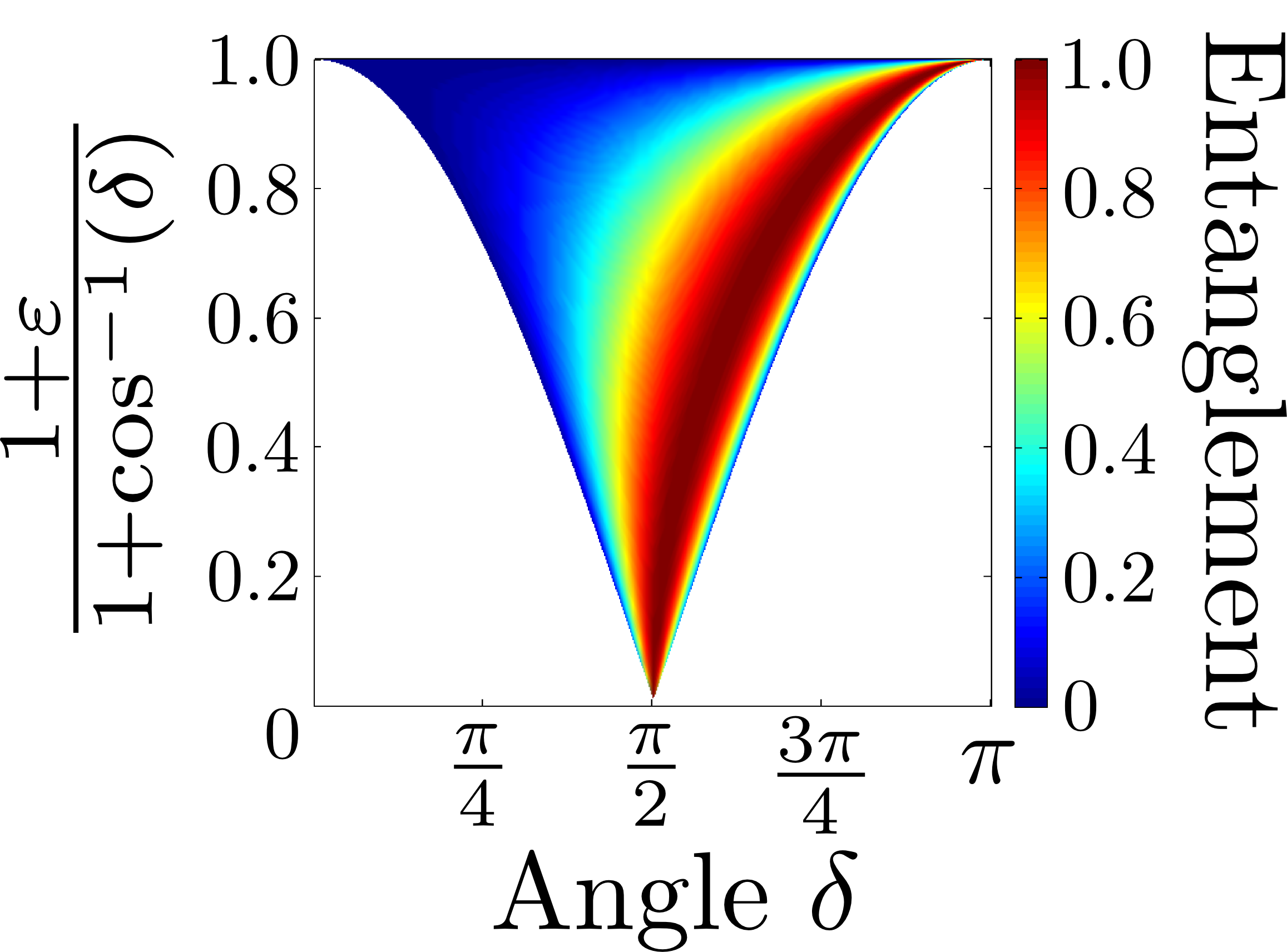}
    }
    \caption{Entanglement creation using a GCNOT. a) Classical states $\{\ket{c_0},\ket{c_1}\}$ 
	(black dashed arrows). 
	The states $\{\ket{0},\ket{1}\}$ (red dotted arrows) are nontrivial superpositions of those basis states. b) Output entanglement after applying a GCNOT $\Lambda_{\theta,\varepsilon}$ to the input state $\ket{0}$. With the optimal choice of  $\varepsilon=\varepsilon^\mathrm{opt.}(\theta)$, exactly one ebit of entanglement is created.}
    \label{fig:GCNOT_ent}
\end{figure}

Fortunately, Theorem 2 provides us another option. Conceptually, we switch to a picture
where $\{\ket{0},\ket{1}\}$ are \emph{not} the classical states, e.g., by choosing
\begin{align}
\ket{c_{0/1}}=\cos(\tfrac{\theta}{2})\ket{0}\pm\sin(\tfrac{\theta}{2})\ket{1}
\end{align} 
for some fixed $\theta\in(0,\pi)$ (Fig. \ref{fig:GCNOT_ent}a). 
For this choice, we have
\begin{align}
	\ket{0/1} =  \tfrac{1}{\mathcal{N}_{0/1}}(\ket{c_{0}}\pm\ket{c_{1}}),
\end{align} 
where $\mathcal{N}_{0}=2\cos(\tfrac{\theta}{2})$ and $\mathcal{N}_{1}=2\sin(\tfrac{\theta}{2})$.
The states $\{\ket{0},\ket{1}\}$ are thus nonclassical in this picture for any $\theta\in(0,\pi)$. We now find conversion operations $\Lambda = \Lambda_{\theta,\varepsilon}$ such that, for $i=0,1$,
\begin{align}
	\Lambda_{\theta,\varepsilon}[\ket{c_i}] =& \ket{e_i(\theta,\varepsilon)}\otimes\ket{f_i(\theta,\varepsilon)}\label{eq:GCNOT1}
\end{align} 
where $\braket{e_0}{e_1}=(1+\varepsilon)\braket{c_0}{c_1}=(1+\varepsilon)\cos(\theta)$ and $\braket{f_0}{f_1}=\tfrac{1}{1+\varepsilon}$. We shall refer to this type of transformations as a \emph{generalized CNOT} (GCNOT).

In Fig. \ref{fig:GCNOT_ent}b, we show the entanglement of the output states $\Lambda_{\theta,\varepsilon}[\ket{0}]$ for the possible values of $\theta$ and $\varepsilon$. We see that for any $\tfrac{\pi}{2} \leq \theta < \pi$, there is an optimal choice
$\varepsilon^\mathrm{opt.}(\theta)$ such that $\Lambda_{\theta,\varepsilon^\mathrm{opt.}}[\ket{0}]$ generates a maximally entangled state. If we use the input state $\ket{1}$, the resulting plot is mirrored horizontally. Hence, despite the states $\{\ket{0},\ket{1}\}$ having no resource value from the perspective of coherence, they do have value from the perspective of the choice $\mathcal{C}_P=\{\ket{c_0},\ket{c_1}\}$. Both of these states can be used, with the appropriate conversion operations, to generate maximally entangled states. 

Of course, the operation $(U^\dagger\otimes\iden) [CNOT] (U\otimes\iden)$, with $U=H$ being the Hadamard gate, also transforms the computational basis states into maximally entangled states. With the specified operational constraints, namely that we cannot locally create coherence, such transformations were off-limits for us. Fortunately, except at the special point $\theta=\tfrac{\pi}{2}$, the optimal GCNOTs are not local-unitarily equivalent to the CNOT. To see why, notice that the standard CNOT sends at least \emph{two orthogonal states} into maximally entangled states (e.g., $\ket{\pm}\mapsto\tfrac{1}{\sqrt{2}}[\ket{00}\pm\ket{11}]$). All gates local-unitarily equivalent to a CNOT also have this property. For any $\theta\neq\tfrac{\pi}{2}$, the optimal GCNOT converts \emph{only one} possible input state to a maximally entangled state. For instance, if $\pi/2<\theta<\pi$, the state $\ket{0}$ can be converted to a maximally entangled state. The same conversion operation applied to $\ket{1}$ also yields an entangled state, but it is \emph{not} maximal (cf. Fig. 1b and its mirror image). For $0<\theta<\tfrac{\pi}{2}$, the roles of $\ket{0}$ and $\ket{1}$ are reversed. Only the state which is the \emph{maximal distance} (geometrically on the Bloch sphere) from the classical states is converted into a maximally entangled state. For orthogonal classical states, there are many states at the same distance; in the nonorthogonal setting, there is a unique maximal state. Thus, GCNOTs belong to a different class of entanglement generating gates than regular CNOTS.

\section{S2. Connecting entanglement theory and nonclassicality}

Another door that opens up once we have convertibility between
nonclassicality and entanglement is that we can apply tools and methods from one picture to infer something about the other. One obvious candidate is to create \emph{nonclassicality witnesses} out of entanglement witnesses. That we can even have nonclassicality witnesses follows from the assumed convex structure of the classical states and the Hahn-Banach theorem in the same way as it does for entanglement theory. For settings with the convertibility property, we would expect nonclassicality and entanglement witnesses to be intimately connected. 

As an illustration, suppose we have a set of classical states $\mathcal{C}_P=\{\ket{c_i}\}_{i\in\mathcal{I}}$ which satisfies either Theorem 2 or 3. Let $\mathcal{H}_A=\mathrm{span}(\mathcal{C}_P)$ be the state space for our input system, and $\mathcal{H}_B\cong\mathcal{H}_A$ be for the ancilla system, with reference state $\ket{\psi_\mathrm{ref}}$. Arbitrarily fix some conversion operation $\Lambda$. If $\rho_\mathrm{in}$ is nonclassical, the corresponding output state $\rho_\mathrm{out}=\Lambda(\rho_\mathrm{in}\otimes\ketbra{\psi_\mathrm{ref}}{\psi_\mathrm{ref}})\Lambda^\dagger$ is entangled. Thus, there exists an entanglement witness $W$ that can detect the entanglement of $\rho_\mathrm{out}$, i.e., an observable such that
\begin{align}
	& \tr{W\rho_\mathrm{out}}< 0,\\
	&  \tr{W\ketbra{\psi_A}{\psi_A}\otimes\ketbra{\psi_B}{\psi_B}}\geq  0,
\end{align}
for all $\ket{\psi_{A}}\in\mathcal{H}_{A}$ and $\ket{\psi_{B}}\in\mathcal{H}_{B}$.

Define a new observable $W':=\Lambda^\dagger W\Lambda$. This satisfies 
\begin{align}
	& \tr{W'\rho_\mathrm{in}\otimes\ketbra{\psi_\mathrm{ref}}{\psi_\mathrm{ref}}}<0\\
	& \tr{W'\ketbra{c_i}{c_i}\otimes\ketbra{\psi_\mathrm{ref}}{\psi_\mathrm{ref}}}\geq 0
\end{align}
for all classical states $\{\ket{c_i}\}$. Taking one further step, we restrict $W'$ using the $\ket{\psi_\mathrm{ref}}$ subspace of the ancilla system. This results in an observable on $\mathcal{H}_A$ only:
\begin{align}
	\widetilde{W} := (\iden\otimes\bra{\psi_\mathrm{ref}})W'(\iden\otimes\ket{\psi_\mathrm{ref}}).
\end{align}
From the above relations, we must have
\begin{align}
	& \tr{\widetilde{W}\rho_\mathrm{in}}<0\\
	& \tr{\widetilde{W}\ketbra{c_i}{c_i}}\geq 0
\end{align}
for all classical states $\ket{c_i}$. Hence, an entanglement witness $W$ on the output 
system is converted to a 
nonclassicality witness $\widetilde{W}$ on the input system. The only requirement was that $W$ had to detect the entanglement of $\rho_\mathrm{out}$, a state which is in the image of the transformation $\Lambda$.

A related idea is \emph{entanglement potential} \cite{asboth05a}. Instead of transforming a bipartite witness $W$ backwards into a single-system observable $\widetilde{W}$, we physically carry out the conversion operation $\Lambda$ and examine the entanglement at the output. This has two potential strengths. First, we can use entanglement measures to indirectly quantify nonclassicality. For this, one must first construct a well-defined resource theory for nonclassicality, defining not just the classical/nonclassical states, but also the allowed operations. We are currently investigating this direction. 

The second advantage of transferring to an entanglement-based setting is that it allows us to overcome operational constraints or superselection rules that may be present in the single-system setting. Take as example a system of identical particles in the same mode, each with
total (pseudo)spin $J=\tfrac{1}{2}$. Due to symmetrization, observables which are `local' to each particle are not measurable. However, standard experimental methods allow us to measure the mean values and variances of the collective spin operators $\hat{J}_k$ in various directions. Other collective observables can be much more difficult to measure. Thus, optimal criteria have been found
for detecting nonclassicality with only the means and variances of the $\hat{J}_k$ \cite{toth09a}. These criteria can detect all nonclassical states that are detectible with such information. 

However, the mean and variance information is not sufficient to detect every nonclassical state. We can take advantage of Theorem 3 to overcome this deficiency, converting nonclassicality (i.e., particle entanglement) faithfully into mode entanglement. The algebraic structure of the 
system is exactly the same in both pictures. Thus, we can independently measure the means and variances of the mode spin operators $\hat{J}^{A/B}_k$, and use this information as if we had measured $\hat{J}_k$ on separate clusters containing $N_X$ and $N_Y$ particles in the original single-mode situation. This extra information may allow us to detect more entangled states than we could with collective measurements alone.

Finally, we can reverse the direction, picturing entanglement theory 
(or at least a subset of it) through a local lens. It has already been recognized \cite{streltsov15a,winter15a,chitambar15a,streltsov15b} that the resource theory of coherence bears strong resemblance to entanglement theory on the so-called \emph{maximally-correlated states}. The results presented here reveal further links between entanglement theory and local nonclassicality. 

\section{S3. Linking settings: GCNOTS and beamsplitters}

Because of its generality, our nonclassicality framework also allows us to connect ideas from seemingly distinct settings, with the ultimate hope that this cross-pollination will lead to new physical insights. As illustration, we will explore an interesting connection between discrete systems and the optical setting. Specifically, we will compare the two relevant conversion operations: GCNOTs and beamsplitters. Both of these achieve the same kind of structural transformation, (non)classical $\rightarrow$ (non)separable, but in conceptually different ways. However, as we will show below, a GCNOT transformation can, in a simple scenario, be seen as equivalent to some beamsplitter transformation.

Consider first the optics setting. The standard classical states are the coherent states
\begin{equation}
	\ket{\alpha}=e^{\frac{-|\alpha|^2}{2}} \sum_{k=0}^\infty \frac{\alpha^k}{\sqrt{k!}}\ket{k},
\end{equation}
with overlaps given by
\begin{equation}
	\braket{\alpha}{\beta} = \exp(-\tfrac{1}{2}[|\alpha|^2+|\beta|^2-2\alpha^*\beta]).
\end{equation}
As remarked in the main text (Eq. (4)), these overlaps are amenable to the following exponent factoring trick:
\begin{align}\label{eq:opcohsplit_SM}
	\braket{\alpha}{\beta} 
	& = \braket{\alpha}{\beta}^{|r|^2}\braket{\alpha}{\beta}^{|t|^2}
	= \braket{r\alpha}{r\beta}\braket{t\alpha}{t\beta},
\end{align}
for any parameters $(r,t)$ such that $|r|^2+|t|^2=1$. This splitting
is not dependent on the amplitudes $\alpha,\beta$, and works for all coherent states. Of course, the physical transformation realizing this splitting is none other than a beamsplitter, with parameters $r$ and $t$. Suppose we begin with a Schr\"{o}dinger cat state in a single mode,
\begin{align}\label{eq:psi_in_opt}
	\ket{\psi_\mathrm{in}}=\tfrac{1}{\sqrt{\mathcal{N}}}[\ket{\alpha}+\ket{\beta}],
\end{align}
where $\mathcal{N}:=2+2\mathrm{Re}\braket{\alpha}{\beta}$. For $|\braket{\alpha}{\beta}|\neq 1$, this is a nonclassical state of the optics setting. Using a beamsplitter, we can convert this to a two-mode entangled state,
\begin{align}
	\Lambda^{BS}\ket{\psi_\mathrm{in}}=\tfrac{1}{\sqrt{\mathcal{N}}}[\ket{r\alpha}\ket{t\alpha}
	+\ket{r\beta}\ket{t\beta}].
\end{align}

Notice now that the two states $\{\ket{\alpha},\ket{\beta}\}$ are linearly independent. Therefore they also satisfy the conditions of Theorem 2. This means that Eq. (\ref{eq:psi_in_opt}) is also a nonclassical state in a specific discrete scenario, and we can identify some GCNOT transformation $\Lambda^{GCNOT}_\varepsilon$, parameterized by some valid overlap scaling $\varepsilon\geq 0$ (note that we allow $\varepsilon=0$ for this specific example). This will give
\begin{equation}
	\Lambda^{GCNOT}_\varepsilon\ket{\psi_\mathrm{in}}=\tfrac{1}{\sqrt{\mathcal{N}}}[\ket{e_\alpha}\ket{f_\alpha}
	+\ket{e_\beta}\ket{f_\beta}],
\end{equation}
where 
\begin{align}
	\braket{e_\alpha}{e_\beta} = & (1+\varepsilon)\braket{\alpha}{\beta},\label{eq:ee_ovlap}\\
	\braket{f_\alpha}{f_\beta} = & (1+\varepsilon)^{-1}.
\end{align}
Thus, the beamsplitter and the GCNOT provide two options for creating entanglement from $\ket{\psi_\mathrm{in}}$.  

Beamsplitters are, of course, readily available in experiments, whereas GCNOT gates could potentially be much harder to implement. How does the GCNOT gate compare to the beamsplitter? For simplicity, we will now assume that the overlap $\braket{\alpha}{\beta}$ is real. We can always introduce new paramters $x,y\in\mathbb{R}$ in the following way:
\begin{align}
	\braket{\alpha}{\beta}^{x} = & (1+\varepsilon)\braket{\alpha}{\beta}, \\
	\braket{\alpha}{\beta}^y = & (1+\varepsilon)^{-1}.
\end{align}
Since $\braket{\alpha}{\beta}\neq 0$ for all $\ket{\alpha},\ket{\beta}$, we can solve for $x$ and $y$:
\begin{align}
	x = & 1-\frac{\log[(1+\varepsilon)^{-1}]}{\log[\braket{\alpha}{\beta}]},\label{eq:xdef}\\
	y = & \frac{\log[(1+\varepsilon)^{-1}]}{\log[\braket{\alpha}{\beta}]}.\label{eq:ydef}
\end{align}
For $\varepsilon\geq 0$, this forces $x \leq 1$, $y \geq 0$, and $x+y=1$.

Our choice of $\varepsilon$ is restricted by the requirement that the overlap in Eq. (\ref{eq:ee_ovlap}) is not greater than unity, i.e., 
\begin{align}
	(1+\varepsilon)^{-1}\geq\braket{\alpha}{\beta}.
\end{align}
Translating this condition to $x$ and $y$, we must have $x \geq 0$ and $y \leq 1$. Thus, we can make the identifications 
\begin{align}
	x=|r|^2,  & ~ y=|t|^2
\end{align}
with $|r|^2,|t|^2 \leq 1$. So the GCNOT with parameter $\varepsilon$ is equivalent to a beamsplitter with reflection/transmission amplitudes ($|r|^2,|t|^2$) given by Eqs. (\ref{eq:xdef})-(\ref{eq:ydef})!

Now, any finite set of coherent states $\{\ket{\alpha_j}\}_{j=1}^{D}$ is linearly independent \cite{sperling10a}, so the alternative GCNOT approach is applicable for all optical states of the form
\begin{align}\label{eq:psi_in_opt_finite}
	\ket{\psi_\mathrm{in}}\sim\sum_{j=1}^D\ket{\alpha_j},
\end{align}
and for arbitrary $1\leq D<\infty$.
It would be interesting to study how GCNOTs and beamsplitters might be related 
in general:
given a set of $D>2$ linearly independent classical states and a GCNOT, can we always find an equivalent realization using coherent states and a beamsplitter, and vice versa? Or is there some fundamental distinction between GCNOTs and beamsplitters that only appears beyond the two-state case?

\section{S4. An alternate generalization of coherent states}

In this section, we briefly explore an alternative way to generalize optical coherent states in infinite dimensions. This approach can also give splitting properties similar to the classical states considered in the main text. In infinite-dimensional systems the relevant shifting operators can be noncompact, as is the usual situation in quantum optics for $a$ and $a^{\dagger}$, where $[a,a^{\dagger}]=\iden$. 
In this case, instead of the group-theoretic approach one can generalize coherent states as the eigenstates of the lowering operator at hand.
One common example arising in different physical situations \cite{london26a,london27a,susskind64a,deoliveira03a} is the use of bare raising and lowering operators, 
also called exponential phase operators, or Susskind-Glogower operators. 
The bare raising and lowering operators are given respectively by
$
l_- = \sum_{n=0}^{\infty}|n\rangle\langle n+1|$, $l_+ = \sum_{n=0}^{\infty}|n+1\rangle\langle n|$, 
with commutation relation $[l_-,l_+]=|0\rangle\langle 0|$. The eigenstates of $l_-$ are the so-called phase-coherent states \cite{lerner70a}
$
|\mu\rangle = \sqrt{1-|\mu|^2}\sum_{n=0}^{\infty}\mu^n|n\rangle 
$,
so that $l_-|\mu\rangle=\mu|\mu\rangle$.  
Another example are coherent states constructed as eigenstates of the lowering operator of the $SU(1,1)$ algebra. 
In general, the eigenstates $|z\rangle$ of the lowering operator $L_-$ of an infinite-dimensional system, i.e., $L_-|z\rangle=z|z\rangle$ are called Barut-Ghirardello coherent states (BGCS) \cite{barut71a}.

Given two spatially-separated modes $A$ and $B$, define the Hamiltonian
$
H =  \xi L_+^AL_-^B + \xi^*L_-^AL_+^B
$
which, roughly speaking, splits the modes by destroying a quanta in one mode while creating a quanta in the other mode. 
We show now that, in an appropriate regime, the single-mode BGCS do not generate entanglement via the global unitary $U=e^{i\alpha H}$. 
First, we recall the Zassenhaus formula:
\begin{align}
e^{\alpha (M+N)}= e^{\alpha M}e^{\alpha N}e^{-(\alpha^2/2)([M,N]) }(h.o.),
\end{align}
with $(h.o.)$ denoting higher order commutator terms. 
We see that if $[M,N]=\iden$, then $e^{\alpha (M+N)}\propto e^{\alpha M}e^{\alpha N}$. Similarly, whenever $|\alpha^2/2|\ll ||[M,N]||$, then $e^{\alpha (M+N)}\approx e^{\alpha M}e^{\alpha N}$. 
In these situations, we have
\begin{align}
e^{i\alpha H}|z\rangle |0\rangle = & e^{i\alpha\xi (L_-^AL_+^B)}e^{i\alpha\xi^* (L_+^AL_-^B)}|z\rangle|0\rangle \nonumber\\
= & e^{i\alpha\xi  (L_-^AL_+^B)}|z\rangle|0\rangle\nonumber\\
= & |z\rangle\otimes(e^{iz\alpha\xi L_+^B}|0\rangle).
\end{align}
Notice that we only used the property $L_-|z\rangle=z|z\rangle$, i.e., 
the actual form of the BGCS $|z\rangle$ is not relevant, even though the commutation relation between creation and annihilation operators is.
In the case of phase-coherent states, for example, $[l_-,l_+]=|0\rangle\langle 0|$ and thus we have the interaction regime 
$
|i\alpha^2/2|\ll ||[l_-^Al_+^B,l_+^Al_-^B]|| = ||(|0_A\rangle\langle 0_A|-|0_B\rangle\langle 0_B|)||
$.

\bibliography{non-classical-refs}



\end{document}